\documentclass{article}

\usepackage{arxiv}
\RequirePackage{epsfig}
\RequirePackage{amssymb}
\RequirePackage{natbib}
\RequirePackage{graphicx}
\RequirePackage{amsthm}

\usepackage[utf8]{inputenc} % allow utf-8 input
\usepackage[T1]{fontenc}    % use 8-bit T1 fonts
\usepackage{hyperref}       % hyperlinks
\usepackage{url}            % simple URL typesetting
\usepackage{booktabs}       % professional-quality tables
\usepackage{amsfonts}       % blackboard math symbols
\usepackage{nicefrac}       % compact symbols for 1/2, etc.
\usepackage{microtype}      % microtypography
\usepackage{lipsum}
\usepackage{graphicx}
\usepackage{mathtools}

\usepackage[table,xcdraw]{xcolor}
\usepackage[caption=false,subrefformat=parens,labelformat=parens]{subfig}
\usepackage{verbatim} 
\usepackage{chemformula}
\usepackage{natbib} 
\usepackage{enumitem}
\usepackage{booktabs}
\usepackage{latexsym}
\usepackage{amsmath}
\usepackage{amsfonts}
\usepackage{amssymb}
\usepackage{amsbsy}
\usepackage{amsthm}
\usepackage{bbm}
\usepackage{bm}
\usepackage{caption}
\usepackage{subcaption}
\usepackage{multirow}
\usepackage{siunitx}
\usepackage[ruled]{algorithm2e}
\usepackage{wrapfig}
\usepackage[title,page]{appendix}

\usepackage{soul}

\newtheoremstyle{boldremark}
    {\dimexpr\topsep/2\relax} % space above
    {\dimexpr\topsep/2\relax} % space below
    {}          % body font
    {}          % indent amount
    {\bfseries} % theorem head font
    {.}         % punctuation after theorem head
    {.5em}      % space after theorem head
    {}          % theorem hed spec. (empty = "normal")

\newtheorem{theorem}{Theorem}

\theoremstyle{boldremark}

\title{Adjoint Sensitivity Analysis on Multi-Scale bioprocess stochastic reaction network}

\author{Keilung Choy  \\
 Northeastern University
 \And
  Wei Xie \thanks{Corresponding author. Email: w.xie@northeastern.edu} \\
 Northeastern University
 }

\date{}
\renewcommand{\shorttitle}{\textbf{Choy and Xie:} \textit{Adjoint Sensitivity Analysis on Multi-Scale bioprocess stochastic reaction network}}

\begin{document}

\maketitle
\begin{abstract}
Motivated by the pressing challenges in the digital twin development for biomanufacturing systems, 
we introduce an adjoint sensitivity analysis (SA) approach to expedite the learning of mechanistic model parameters. 
In this paper, we consider enzymatic stochastic reaction networks representing a multi-scale bioprocess mechanistic model that allows us to integrate disparate data from diverse production processes and leverage the information from existing macro-kinetic and genome-scale models. To support forward prediction and backward reasoning, we develop a convergent adjoint SA algorithm studying how the perturbations of model parameters and inputs (e.g., initial state) propagate through enzymatic reaction networks and impact on output trajectory predictions. 
This SA can provide a sample efficient and interpretable way to assess the sensitivities between inputs and outputs accounting for their causal dependencies.
Our empirical study underscores the resilience of these sensitivities and illuminates a deeper comprehension of the regulatory mechanisms behind bioprocess through sensitivities.
\end{abstract}

% keywords can be removed

\section{INTRODUCTION}
\label{sec:intro}
%\textbf{Problem description.} 
\begin{comment}
To support forward prediction and backward reasoning for biomanufacturing processes, in this paper, we aim to develop an interpretable and sample efficient sensitivity analysis (SA) approach on multi-scale mechanistic models of  enzymatic stochastic reaction networks that characterizes the intrinsic causal relationships from molecular- to macro-kinetics. This mechanistic model is primarily formulated as 
%using continuous-time Markov chains (CTMC) and 
stochastic differential equations (SDEs) representing nonlinear dynamics and state-dependent variations driven by regulatory mechanisms.     
\end{comment}
With prior knowledge of the reaction network structure and regulatory mechanisms, the trajectory output of bioprocess hinges on three key inputs: (1) initial states; (2) actions; and (3) parameters of the mechanistic model\citep{BO_sens}. Suppose the effect of actions, such as feeding strategies, on state change is immediate and known. %observable. 
{\textit{Therefore, our objective is to develop an interpretable and sample-efficient sensitivity analysis (SA) approach for the multi-scale bioprocess mechanistic model that focuses on investigating the sensitivities between inputs (i.e., initial states and model parameters) and outputs.}}

%\textbf{Multi-Scale Bioprocess Mechanistic Modeling.} 
{A multi-scale} bioprocess mechanistic model can facilitate the development of digital twins and BioFoundries for biomanufacturing {processes}. Within this paradigm, the model's foundation is constructed upon fundamental building blocks, i.e., molecular reaction networks.  %are characterized by model parameters with interpretable meaning. 
%This multi-scale mechanistic model can enable a consistent synthesis of heterogeneous data derived from various production processes and leverage on existing macro-kinetic and genome-scale models. 
There exist various challenges to conduct SA on the mechanistic model of enzymatic stochastic reaction networks. 
One significant feature 
%of the enzymatic stochastic reaction network 
is its double-stochasticity, which means at any time molecular reaction rates are contingent upon random states, such as species concentrations and environmental variables.
%while they are rendered as random variables with reaction rates serving as parameters.
Upon formulating the multi-scale bioprocess mechanistic model in the form of stochastic differential equations (SDEs), the drift and diffusion terms could be built based on Michaelis–Menten kinetics, which is the most frequently used kinetic model of enzymatic reactions found in existing literature \citep{Sarantos2018}. It leverages our existing understanding of the reaction rate structure and encapsulates the double stochasticity inherent in stochastic reaction networks, i.e., both the drift and diffusion terms are contingent upon the current states. Consequently, deriving analytical solutions for such SDEs, characterizing stochastic molecular reaction network dynamics and variations, could be difficult. Moreover, the interactions between input factors and intermediate states further
%influence output dynamics and variability, which 
introduce high complexity in SA.
\begin{comment}
Therefore, in this paper, we aim to leverage the information provided by these interactions, which not only reduces the complexity of the sensitivity analysis but also enables the results to reflect deeper insights within the reaction network.    
\end{comment}

%\textbf{Existing sensitivity analysis approaches.} 
Existing sensitivity analysis approaches could be divided into two categories: local and global sensitivity analysis. Global sensitivity analysis focuses on the impact of significant variations in model inputs. For example, the Sobol method \citep{Sobol2001GlobalSI} is a global sensitivity analysis approach based on variance decomposition, which decomposes the variance of the model output into the contributions of different model inputs and parameters.
%a series of variances across different combinations of input parameters.
In contrast, the objective of local sensitivity analysis is to quantify the effects of minor perturbations in model inputs and parameters on model predictions. 
%We mentioned that in the Multi-Scale Bioprocess Model, parameters explain real-life bioprocess and have real physical meanings, which means we do not expect large variations in parameters. Thus, we select the local sensitivity analysis approach for our study.
There are a variety of local sensitivity analysis approaches including finite direct differential method \citep{senschem}, Nominal Range Sensitivity Method \citep{PTEA}, and automatic differentiation \citep{adcp}. Unfortunately, most of these existing approaches are not applicable to our study since they scale poorly, either in terms of computational time or memory usage, as the number of parameters and states within the model increases. Given that the multi-scale bioprocess model typically represents a complex system with many inputs and parameters, these limitations pose a significant barrier to the learning process of underlying mechanisms.

\textit{In this paper, we formulate the multi-scale bioprocess mechanistic model in SDEs form, accounting for underlying causal interdependencies of an enzymatic stochastic reaction network, and then develop an adjoint SA approach studying the sensitivities between inputs and outputs}. It can correctly and efficiently quantify the contribution and criticality of each input and model parameter impacting on the prediction errors of multivariate output trajectories, such as productivity and product critical quality attributes (CQAs). %, and efficiently predict their local sensitivities.  
To support forward prediction and backward reasoning, this adjoint sensitivity analysis over the operator of SDEs, characterizing bioprocess mechanisms, exhibits robust scalability even when the complexity of the mechanistic model increases. In addition, since it leverages the structural information of the regulatory reaction network, the adjoint SA can provide sample efficient and interpretable guidance to search inputs and model parameters, accounting for their interactions, to speed up the learning process.
%between states and parameters.

\begin{comment}
This paper makes two contributions to the field. Firstly, it addresses a prevalent challenge in bioprocess modeling, that is, the parameters of mechanistic models are often estimated from limited real-world data, resulting in the estimation errors. The proposed local SA approach is not only beneficial to the expedition of process mechanism learning but also facilitates a comprehensive assessment of the influence of model parameters on the prediction accuracy. From this perspective, this study successfully identifies the bottlenecks of bioprocess mechanistic model and
%and clarifies the roles these factors are playing in the regulatory mechanisms. 
it could advance the understanding of enzymatic reaction network mechanisms, speed up digital twin development, and guide more efficient design of experiments. This plays a critical role since biomanufacturing is a highly regulated industry and each experiment is expensive. Secondly, the paper addresses the challenge of handling the intricate interdependencies inherent in bioprocesses. By leveraging structural information within the enzymatic reaction networks, the paper mitigates the complexity introduced by the key properties such as double stochasticity in bioprocess mechanistic models and further streamlines the calculation of input-output sensitivities, which alleviates the computational bottleneck encountered when conducting local SA in previous studies.
\end{comment} 

{
This paper addresses two key challenges in bioprocess modeling through the development of the adjoint SA approach. Firstly, it addresses the issue that the dimensionality of model parameters can be so high that searching for the optimal parameter estimation becomes challenging. % due to limited real-world data. 
The local SA approach assesses the impact of model parameters on prediction accuracy and expedites process mechanism learning. This facilitates digital twin development and more efficient experimental design, which is crucial in the costly, highly regulated biomanufacturing industry. Secondly, the proposed adjoint SA can incorporate the intricate interdependencies of bioprocesses. By leveraging structural information within enzymatic molecular reaction networks, the paper mitigates complexity and streamlines the calculation of input-output sensitivities, alleviating computational bottlenecks encountered in previous local SA studies.
}

The structure of the paper is organized as follows. In Section~\ref{sec:problemDescription}, we initiate by delineating the multi-scale bioprocess mechanistic model and summarizing its key characteristics. Subsequently, in Section~\ref{sec:Meta}, we proceed to construct a metamodel to the multi-scale bioprocess, leveraging the insights from the preceding section. Section~\ref{sec:LS-multiScaleModel} is dedicated to local sensitivity analysis algorithm development, which will help us investigate the barriers to the reduction of model prediction errors and provide a guidance to speed up the search for optimal inputs.
%for optimal data acquisition. 
Building upon this part, Section~\ref{sec:algorithm} introduces an adjoint SA algorithm on SDEs designed to enhance computational efficiency. Improvements brought by the algorithm will be validated in Section~\ref{sec: empirical study} through the empirical study of its finite-sample performance. Finally, in Section~\ref{sec: conclusion}, we synthesize the findings and insights gathered throughout this study, leading to the paper's conclusion.

\section{Problem Description}
\label{sec:problemDescription}
\begin{comment}
    
In Section~\ref{sec:multi-scaleBioprocess}, we first describe multi-scale bioprocess mechanistic model characterizing the mechanism of biomanufacturing process and causal interdependence from molecular- to macro-scope kinetics. It is built on a fundamental stochastic molecular reaction network with reaction rate as a function of the current state, including species concentration and environmental conditions (such as pH and temperature). This multi-scale mechanistic model can support the digital twin development. Then, we discuss the key properties and challenges on the multi-scale mechanistic model, including double stochasticity. 
In this paper, suppose the model structure of the regulatory molecular reaction network is known. % and it is built on biological/physical/chemical mechanisms. 
Then, a local sensitivity analysis is considered to facilitate the most informative data collection for model development. 
\end{comment}

\subsection{Multi-Scale Bioprocess Stochastic Reaction Network 
%Macro-kinetic State Transition Modeling
}
\label{sec:multi-scaleBioprocess}

A multi-scale bioprocessing mechanistic model characterizes the causal dependence from molecular to macroscopic kinetics and it is built on the fundamental building block, i.e., enzymatic molecular reaction networks. Suppose the system is composed of $I$ species, denoted by $\pmb{X}=(X_1, X_2, \ldots, X_I)^\top$, interacting with each other through $J$ reactions. 
At any time $t$, let $\pmb{s}_t = (s_t^1,s_t^2,\ldots,s_t^{I})^\top \in \mathbb{R}_+^{I}$ be the bioprocess state, where $s_t^i$ denotes the number of molecules of species $i$. Each $j$-th reaction with $j=1,2,\ldots,J$ is characterized by a reaction vector $\pmb{N}_j \in \mathbb{R}^{I}$, describing the change in the numbers of $I$ species' molecules when a $j$-th molecular reaction occurs. The associated reaction rate denoted by $v_j$, depending on state, such as the current number of molecules of each species, describes the rate at which the $j$-th reaction occurs. Specifically, for the $j$-th reaction equation given by
\begin{equation*}
    p_{j1} X_1 + p_{j2} X_2 + \cdots + p_{jI} X_{I} \xrightarrow{\bm{v}_j} q_{j1} X_1 + q_{j2} X_2 + \cdots + q_{jI} X_{I},
\end{equation*}
the reaction relational structure, specified by the vector $\pmb{N}_j = (q_{j1}-p_{j1},q_{j2}-p_{j2},\ldots,q_{jI}-p_{iI})^\top$, is known for $j=1,2,\ldots,J$.
Thus, the stoichiometry matrix $\pmb{N} = \left(\pmb{N}_1, \pmb{N}_2, \ldots, \pmb{N}_{J}\right) \in \mathbb{R}^{I \times J}$ characterizes the structure information of the reaction network composed of $J$ reactions. The $(i,j)$-th element of $\pmb{N}$, denoted by $N_{ij}$, represents the number of molecules of the $i$-th species that are either consumed (indicated by a negative value) or produced (indicated by a positive value) in each random occurrence of the $j$-th reaction.

In this paper, we suppose that the structure of the reaction network represented by matrix $\pmb{N}$ is known. The regulation mechanism of each $j$-th reaction is characterized by the reaction rate function $\bm{v}_{j}$, which is associated with the current system state $\pmb{s}_{t}$ and the mechanistic model parameters denoted by $\pmb{\theta}_t$. 
Let $\bm{R}_{t}$ %$ \in \mathbb{R}_{+}^{R}$ 
be a vector of the occurrences of each molecular reaction in a given short time interval $(t,t +\Delta t]$ and the system state is updated from $\pmb{s}_{t}$ to $\pmb{s}_{t+1}$. 
Since a molecular reaction will occur when one molecule {collides, binds, and reacts} with another one while molecules move around randomly, driven
by stochastic thermodynamics of Brownian motion \cite{golightly2005bayesian},
the occurrences of molecular reactions %during a short interval $(t,t+\Delta t]$ 
are modeled by non-homogeneous Poisson process.
%We assume that two reactions cannot occur at exactly the same time and $\bm{R}_{t}$ %is a counting process. Then, we represent $R_{t}$ in terms of the most elementary counting process, namely, the follows a Poisson process. 
Therefore, the state transition model becomes, 
%characterizing stochastic reaction network macro-kinetics can be modeled as:
%$\bm{N}$ is the stoichiometry matrix and 
\begin{equation}
\pmb{s}_{t+1}=\pmb{s}_{t}+\bm{N}\cdot\bm{R_{t}}
\quad \mbox{with} \quad \bm{R}_{t}\sim \mbox{Poisson}(\bm{v}(\pmb{s}_{t},\bm{\theta}_t)), \label{Poisson}
\end{equation}
where $\bm{N}\cdot\bm{R_{t}}$ represents the net amount of reaction outputs during time interval $(t,t+\Delta t]$. 

Michaelis–Menten (MM) kinetics is commonly used to model the regulation mechanisms of enzymatic reaction networks and the flux rates $\bm{v}(\pmb{s}_{t},\bm{\theta}_t)$ depend on the state \citep{MM2007}. 
In an enzymatic molecular reaction as shown in the first equation in (\ref{MM}), the substrate (S) initially forms a reversible complex (ES) with the enzyme (E), i.e., the enzyme and substrate have to interact for the enzyme to be able to perform its catalytic function to produce the product (P), 
%Its standard expression is shown in the first equation in (\ref{MM}) with 
with $K_F$, $K_R$, and $K_{cat}$ representing kinetic rates.
%where their product with the corresponding substrate concentrations will be the reaction rates.
%Michaelis–Menten equation is based on (\ref{MM}) and has certain assumptions. First, we assume that during the reaction an equilibrium condition is established for the binding and dissociation of the Enzyme and Substrate. 
For MM kinetics as shown in the second equation in (\ref{MM}), we assume the enzyme is either present as the free enzyme or as the ES complex, i.e., $[E]_{total}=[E]+[ES]$ with $[E]$ denoting the concentration of the enzyme. Suppose the rate of formation of the ES complex is equal to the rate of dissociation plus the breakdown, 
i.e., $K_F[E][S] = [ES](K_R+K_{cat})$.
%Based on the assumptions, we could derive Michaelis–Menten equation, which is the second equation in (\ref{MM}). 
Thus, \textit{the parameters in MM kinetics characterize the regulation mechanisms of enzymatic reaction network}: (1) $V_{max}^{j}=K_{cat}[E]_{total}$, is the maximum possible velocity of the $j$-th molecular reaction that can occur when all the enzyme molecules are bound with the substrate, i.e.,$[E]_{total}=[ES]$; and (2) $K_{m}^{j}=\frac{K_R+K_{cat}}{K_F}=\frac{[E][S]}{[ES]}$ is a dissociation constant for the ES complex,
%and a low value of $K_{m}^{j}$ is often interpreted as a high affinity of the enzyme for the substrate,
\begin{equation}
E+S\underset{K_R}{\overset{K_F}{\rightleftarrows}} ES \overset{K_{cat}}{\rightarrow} E+P \text{(product)} ~~~
\mbox{and} ~~~ \bm{v}_{j}(\pmb{s}_{t},\bm{\theta}_t)= \frac{V_{max}^{j}s_t^{j}}{K_{m}^{j}+s_t^{j}}. \label{MM}
\end{equation}

%Then, we describe the state transition model for bioprocess and use $\pmb{s}_t=\left(s^1_t,...,s^{n_s}_t\right)$ representing the state (i.e., number or concentration of $n_s$ species) and $\pmb{a}_t =\left(a^1_t,...,a^m_t\right)$ representing the action at time $t$, where $1\le t \le H$ with $H$ representing the planning horizon. 

%To introduce our sensitivity analysis approach to the state transition model, we apply a numerical approximation scheme with a fixed time step $\Delta t>0$ based on state transition kinetics shown in Equation (\ref{Poisson}), the model for the discrete-time observations of the continuous-time stochastic process then becomes:

By applying diffusion approximation, the state transition model becomes,
\begin{equation} \label{EMapprox}
p(\pmb{s}_{t+1}|\pmb{s}_{t},\bm{\theta}_t)\sim\left\{
\begin{array}{l}
         \mathcal{N}(\pmb{s}_{t}+\bm{N}\pmb{v}(\pmb{s}_{t},\bm{\theta}_t)\Delta t,\bm{N}diag(\pmb{v}(\pmb{s}_{t},\bm{\theta}_t))\bm{N}^{\top}\Delta t),\pmb{s}_{t+1}>\bm{0},\\
         0,\pmb{s}_{t+1}\leq \bm{0}.
\end{array}
\right.
\end{equation}
%Derived from Equation (\ref{EMapprox}), 
Thus, the multi-scale bioprocess mechanistic model is further represented as a system of Stochastic Differential Equations (SDEs), intricately reliant on parameters $\bm{\theta}_t$ and current states $\pmb{s}_t$. The diffusion term is necessary due to the inherent stochasticity of molecular reactions in the bioprocess:
\begin{eqnarray}
(d\pmb{s}_t,d\bm{\theta}_t)^{\top}
&=&(\mu(\pmb{s}_t,\bm{\theta}_t),f(\pmb{s}_t,\bm{\theta}_t))^{\top}dt+
    (\sigma(\pmb{s}_t,\bm{\theta}_t),
    g(\pmb{s}_t,\bm{\theta}_t))^{\top}dW_t. 
    %(\pmb{s}_{0},\bm{\theta}_{0})^{\top} &=&(\pmb{s},\bm{\theta})^{\top}.
    \label{SDE_state}
\end{eqnarray}
where $\mu$, $f$ are the drift terms and $\sigma$, $g$ are the diffusion terms defined as: 
\begin{eqnarray*}
(\mu(\pmb{s}_t,\bm{\theta}_t),f(\pmb{s}_t,\bm{\theta}_t))^{\top}&=&\mbox{E}[(\pmb{s}_{t+\Delta t},\bm{\theta}_{t+\Delta t})^{\top}-(\pmb{s}_{t},\bm{\theta}_{t})^{\top}|\pmb{s}_{t},\bm{\theta}_{t}],\\
(\sigma^{2}(\pmb{s}_t,\bm{\theta}_t),g^2(\pmb{s}_t,\bm{\theta}_t))^{\top}&=&\mbox{Var}[(\pmb{s}_{t+\Delta t},\bm{\theta}_{t+\Delta t})^{\top}|\pmb{s}_{t},\bm{\theta}_{t}].
\end{eqnarray*}
The specific form of infinitesimal mean and variance on the right side will be derived in Section~\ref{sec:Meta}. This SDE system is the foundational groundwork for the subsequent local sensitivity analysis we will conduct. 
In this paper, $\bm{\theta}_t$ remains constant, resulting in both $f$ and $g$ becoming 0. However, given that we model the dynamics of $\bm{\theta}_t$ in SDE form as shown in Equation (\ref{SDE_state}), our approach can be readily extended to accommodate the scenarios where $\bm{\theta}_t$ varies over time.

Given any feasible policy with action, i.e., $\pmb{a}_t=\pi(\pmb{s}_t)$ at any decision time $t$, we consider the state transition model characterizing the process mechanistic dynamics and inherent stochasticity, i.e., 
\begin{equation*}
    \pmb{s}_{t+1} %|\pmb{s}_t, \pmb{a}_t;\pmb{\theta}_t 
    \sim p(\pmb{s}_{t+1}| \pmb{s}_t, \pmb{a}_t; \pmb{\theta}). 
\end{equation*}
%At the decision time $t$, 
For the biomanufacturing process, suppose the impact of decision $\pmb{a}_t$ (e.g., feeding strategy) on the state is known and happens immediately;
that means %its impact happens much faster than the bioprocess inherent dynamics and given any feasible action $\pmb{a}_t$, 
we get the post-decision state denoted by $\pmb{s}_t^\prime = \pmb{f}(\pmb{s}_t,\pmb{a}_t)$ with a known function $\pmb{f}$.
%For example, in the cell culture process, we consider action ${a}_t$ as media exchange and its impact on the solution concentration in bioreactor leads to the post-decision state $\pmb{s}^\prime_t = \pmb{s}_t \cdot (1- a_t)+ \pmb{s}_{in} \cdot a_t$, where $\pmb{s}_t$ and $\pmb{s}^\prime_t$ are the media composition/concentration in the bioreactor before and after the action, $a_t$ is the percentage of media exchanged, and $\pmb{s}_{in}$ represents the composition/concentration of the added media (suppose it is given). 
%Then, at time $t$, the regulation mechanism is characterized by the reaction rates as the function of post-decision state, denoted by $\pmb{v}(\pmb{s}_t^\prime;\pmb{\theta})$ specified by unknown parameters $\pmb{\theta}$.
For notation simplification, we ignore the impact of action. The proposed sensitivity analysis over inputs (i.e., states and model parameters) is extendable to account for the policy effect.

%Now we formally define control policies and the optimization problem solved in this work. 

\subsection{Parameter selection and calibration}
\label{subsec:MU&SA}

    Local sensitivity analysis studies the changes in the model
prediction outputs with respect to initial input values, i.e., 
%initial states 
$(\pmb{s}_0,%parameter set 
\bm{\theta}_0)$. 
The variations around this local point 
%in the parameter space, which 
are quantified by the sensitivity coefficients \citep{Zi2011SensitivityAA} 
%$^{\top}$.
In this way, we could generate the forward flow $\Phi_{0,T}(\pmb{s},\bm{\theta})$ and derive both $\frac{\partial \Phi_{0,T}(\pmb{s},\bm{\theta})}{\partial \pmb{s}_{0}}$ and $\frac{\partial \Phi_{0,T}(\pmb{s},\bm{\theta})}{\partial \bm{\theta}_{0}}$, then $\frac{\partial \mathcal{L}(\Phi_{0,T}(\pmb{s},\bm{\theta}))}{\partial s}$ and $\frac{\partial \mathcal{L}(\Phi_{0,T}(\pmb{s},\bm{\theta}))}{\partial \bm{\theta}}$. 
%The sensitivity coefficients 
that are intimately connected to the solution of the SDE system (\ref{SDE_state}). In Section~\ref{sec:Meta}, we will derive $\mu(\pmb{s}_t,\bm{\theta}_t)$ and $\sigma(\pmb{s}_t,\bm{\theta}_t)$, showing that both functions are characterized by infinite differentiability, with their first-order derivatives being bounded, i.e., $\mu,\sigma \in C_b^{\infty,1}$. Consequently, once the initial values are given, a unique solution to the system (\ref{SDE_state}) is guaranteed to exist. We use $\Phi_{0,t_2}(\pmb{s}_0,\bm{\theta}_0)$ to represent the solution of the SDEs 
%initial value problem 
in (\ref{SDE_state}) at time $t_2$ and it is 
%$\Phi_{0,t_2}(\pmb{s},\bm{\theta})$ is also 
called forward flow satisfying the property:
\begin{equation*}
     \Phi_{0,t_2}(\pmb{s}_0,\bm{\theta}_0)=\Phi_{0,t}(\Phi_{t,t_2}(\pmb{s}_0,\bm{\theta}_0)) ~~ \mbox{ for } ~~ 0\leq t\leq t_2.
\end{equation*}
To simulate $\Phi_{0,t_2}(\pmb{s}_0,\bm{\theta}_0)$, we consider its calculus form based on Equation (\ref{SDE_state}):
\begin{eqnarray}
    \Phi_{0,t_2}(\pmb{s}_0,\bm{\theta}_0) &=&
    (\pmb{s}_0,\bm{\theta}_0)^{\top}+\int_{0}^{t_2}(\mu(\Phi_{0,t}(\pmb{s}_0,\bm{\theta}_0)),f(\Phi_{0,t}(\pmb{s}_0,\bm{\theta}_0)))^{\top}dt \nonumber \\ 
    && +\int_{0}^{t_2}(\sigma(\Phi_{0,t}(\pmb{s}_0,\bm{\theta}_0)),g(\Phi_{0,t}(\pmb{s}_0,\bm{\theta}_0)))^{\top}\circ dW_t,\label{forward_calc}
\end{eqnarray}
where $\circ dW_t$ represents the Stratonovich stochastic integral. For a continuous semimartingale $\{f_t\}_{t<T}$ adapted to the forward filtration $\{\mathcal{F}_{0,t}\}_{t<T}$, the Stratonovich stochastic integral is:
    \begin{equation*}
        \int_0^T f_t \circ dW_t=\lim_{|\Pi| \to 0}\sum\limits_{i=1}^{N}\frac{(f_{t_{i-1}}+f_{t_i})}{2}(W_{t_i}-W_{t_{i-1}}),
    \end{equation*}
where $\Pi=\{0=t_0<0<\ldots<t_N=T\}$ is a partition of time interval $[0,T]$ and $|\Pi|=\max\limits_{n} (t_n-t_{n-1})$. The reason to introduce Stratonovich stochastic integral is that we could generate the inverse flow $\psi_{0,t_2}=\Phi_{0,t_2}^{-1}$ from SDE system (\ref{SDE_state}) based on Equation (\ref{forward_calc}) \cite{kunita2019stochastic}:
\begin{eqnarray}
    \psi_{0,t_2}(\pmb{s}_{t_2},\bm{\theta}_{t_2})&=&(\pmb{s}_{t_2},\bm{\theta}_{t_2})^{\top}-\int_{0}^{t_2}(\mu(\psi_{t,t_2}(\pmb{s}_{t_2},\bm{\theta}_{t_2})),f(\psi_{t,t_2}(\pmb{s}_{t_2},\bm{\theta}_{t_2})))^{\top}dt\nonumber\\
    && - \int_{0}^{t_2}(\sigma(\psi_{t,t_2}(\pmb{s}_{t_2},\bm{\theta}_{t_2})),g(\psi_{t,t_2}(\pmb{s}_{t_2},\bm{\theta}_{t_2})))^{\top}\circ d\widetilde{W}_t,
    \label{backward_calc}
\end{eqnarray}
where $\widetilde{W}_t$ is the backward Wiener process defined as $\widetilde{W}_t=W_t-W_T$ for any $ t<T$. It is adapted to the backward filtration $\{\mathcal{F}_{t,T}\}_{t<T}$. The difference between Equations (\ref{forward_calc}) and (\ref{backward_calc}) is only the negative sign, and such symmetry is attributed to the use of Stratonovich stochastic integral.

We could further define a scalar loss function $\mathcal{L}$ of $\Phi_{0,T}(\pmb{s}_0,\bm{\theta}_0)$. Then, the loss for system (\ref{SDE_state}) becomes $\mathcal{L}(\Phi_{0,T}(\pmb{s}_0,\bm{\theta}_0))$ and the sensitivity coefficient for $\pmb{s}_0$ and $\bm{\theta}_0$ becomes,
\begin{equation}
   A_{0,T}(\pmb{s}_0,\bm{\theta}_0) \equiv  \left(\frac{\partial \mathcal{L}(\Phi_{0,T}(\pmb{s}_0,\bm{\theta}_0))}{\partial \pmb{s}_0},\frac{\partial \mathcal{L}(\Phi_{0,T}(\pmb{s}_0,\bm{\theta}_0))}{\partial \bm{\theta}_0}\right)^{\top}.
   \label{eq.A0T1}
\end{equation}
%which we defined as $A_{0,T}(\pmb{s},\bm{\theta})$. 
Then, based on the chain rule:
\begin{eqnarray}
    A_{0,T}(\pmb{s}_0,\bm{\theta}_0)=\triangledown_{\Phi}\mathcal{L}(\Phi_{0,T}(\pmb{s}_0,\bm{\theta}_0))\triangledown\Phi_{0,T}(\pmb{s}_0,\bm{\theta}_0).
   \label{eq.A0T2}
\end{eqnarray}

For $\triangledown\Phi_{0,T}(\pmb{s}_0,\bm{\theta}_0)$ in (\ref{eq.A0T1}), we first derive $\left(\frac{\partial \pmb{s}_T}{\partial \pmb{s}_0},\frac{\partial \bm{\theta}_T}{\partial \bm{\theta}_0}\right)^{\top}$ under the assumption that $(\pmb{s}_T,\bm{\theta}_T)^{\top}$ is deterministic, i.e., not dependent on the Wiener process $W_t$. We then extend to the case that $(\pmb{s}_T,\bm{\theta}_T)^{\top}$ is stochastically obtained from forward flow $\Phi_{0,T}(\pmb{s}_0,\bm{\theta}_0)$ from SDE system (\ref{SDE_state}) and derive $\left(\frac{\partial \Phi_{0,T}(\pmb{s}_0,\bm{\theta}_0)}{\partial \pmb{s}_0},\frac{\partial \Phi_{0,T}(\pmb{s}_0,\bm{\theta}_0)}{\partial \bm{\theta}_0}\right)^{\top}$, which implements the structural information in the stochastic reaction network through a dual process of forward and backward propagation.

Then, for $\triangledown_{\Phi}\mathcal{L}(\Phi_{0,T}(\pmb{s}_0,\bm{\theta}_0))$ in (\ref{eq.A0T2}), it is determined by the loss function $\mathcal{L}$. Denote the observed bioprocess states as $\pmb{s}^{c}$, the underlying true parameters of the real reaction network as $\bm{\theta^{c}}$ and the output as $O(\pmb{s}^{c})$ which is a scalar function of states. 
Here the superscript ``c" indicates observations and parameters from the real system.
Then the simulated output is denoted by $O(\pmb{s})$, where $\pmb{s}$ is the simulated state with initial state $\pmb{s}_{0}=\pmb{s}^{c}_{0}$ and $\bm{\theta}_0$ as the input of simulation model. $\mathcal{L}$ is a scalar function measuring the difference of observed and simulated outputs, such as mean squared error (MSE) $, i.e., \mbox{E}[(O(\pmb{s}^{c})-O(\pmb{s}))^2|\pmb{s}_0,\bm{\theta}_0]$. Then we could generate $\triangledown_{\Phi}\mathcal{L}(\Phi_{0,T}(\pmb{s}_0))$. Combining the above results, we can derive the sensitivities $ A_{0,T}(\pmb{s}_0,\bm{\theta}_0)$, which can identify the key contributor to the model prediction MSE.

% ===============================

\section{Metamodeling and Prediction}\label{sec:Meta}
%\subsection{Metamodeling}
In this section, we will derive the expression of $\mu$ and $\sigma$, we start with two-step transition from initial states, %i.e., $\mbox{E}(\pmb{s}_{2}|\pmb{s}_{0},\bm{\theta}_0)$,
\begin{eqnarray} 
\mbox{E}(\pmb{s}_{2}|\pmb{s}_{0},\bm{\theta}_0)&=&\int p(\pmb{s}_{1},\bm{\theta}_1|\pmb{s}_{0},\bm{\theta}_0)
\mbox{E}(\pmb{s}_{2}|\pmb{s}_{1},\bm{\theta}_1)d(\pmb{s}_1,\bm{\theta}_1)=\int p(\pmb{s}_{1},\bm{\theta}_1|\pmb{s}_{0},\bm{\theta}_0)[\pmb{s}_{1}+\bm{N}\bm{v}(\pmb{s}_{1},\bm{\theta}_1)\Delta t]d(\pmb{s}_1,\bm{\theta}_1)\nonumber\\
&=&\mbox{E}(\pmb{s}_{1}|\pmb{s}_{0},\bm{\theta}_0)+\int p(\pmb{s}_{1},\bm{\theta}_1|\pmb{s}_{0},\bm{\theta}_0) \bm{N}\bm{v}(\pmb{s}_{1},\bm{\theta}_1)\Delta td(\pmb{s}_1,\bm{\theta}_1).\label{fsexp}
\end{eqnarray}
By combining Equations (\ref{fsexp}) and $\mbox{E}(\pmb{s}_{1}|\pmb{s}_{0},\bm{\theta}_0)=\pmb{s}_{0}+\bm{N}\bm{v}(\pmb{s}_{0},\bm{\theta}_0)\Delta t.\label{PoissonExp}$, we have
\begin{equation}\label{ssexp}
\mbox{E}(\pmb{s}_{2}|\pmb{s}_{0},\bm{\theta}_0)=\pmb{s}_{0}+\bm{N}\bm{v}(\pmb{s}_{0},\bm{\theta}_0)\Delta t+\bm{N}\int p(\pmb{s}_{1},\bm{\theta}_1|\pmb{s}_{0},\bm{\theta}_0)\bm{v}(\pmb{s}_{1},\bm{\theta}_1)\Delta td(\pmb{s}_1,\bm{\theta}_1).
\end{equation}

We need to calculate $\int p(\pmb{s}_{1},\bm{\theta}_1|\pmb{s}_{0},\bm{\theta}_0)\bm{v}(\pmb{s}_{1},\bm{\theta}_1)\Delta td(\pmb{s}_1,\bm{\theta}_1)$ in Equation~(\ref{ssexp}), i.e., the expectation of flux rate after one step transition.
%We consider the MM kinetics in (\ref{MM}). 
For each  $j$-th reaction, its reaction rate ${v_{j}}(\pmb{s}_{1},\bm{\theta}_1)$ is related to $s_{1}^{j}$, and by applying the MM kinetics in (\ref{MM}), we have
%, so we could replace $p(\pmb{s}_{1},\bm{\theta}_1|\pmb{s}_{0},\bm{\theta}_0)$ with marginal distribution of components, which means:
%\begin{flalign}
 \begin{eqnarray}   
 &&\int p(\pmb{s}_{1},\bm{\theta}_1|\pmb{s}_{0},\bm{\theta}_0)\bm{v_{j}}(\pmb{s}_{1},\bm{\theta}_1)\Delta td(\pmb{s}_1,\bm{\theta}_1)=\int p(s_{1}^{j},\bm{\theta}_{1}|\pmb{s}_{0},\bm{\theta}_0)\frac{V_{max}^{j}s_{1}^{j}}{K_{m}^{j}+s_{1}^{j}}\Delta td(\pmb{s}_1,\bm{\theta}_1)
 \nonumber\\
    &=&\int p(s_{1}^{j},\bm{\theta}_{1}|\pmb{s}_{0},\bm{\theta}_0)\left(V_{max}^{j}-\frac{V_{max}^{j}K_{m}^{j}}{K_{m}^{j}+s_{1}^{j}}\right)\Delta td(\pmb{s}_1,\bm{\theta}_1)\nonumber\\
    &=&V_{max}^{j}\Delta t-V_{max}^{j}K_{m}^{j}\Delta t\int p(s_{1}^{j},\bm{\theta}_{1}|\pmb{s}_{0},\bm{\theta}_0)\frac{1}{K_{m}^{j}+s_{1}^{j}}d(\pmb{s}_1,\bm{\theta}_1).
    \label{ssexpMMin}
\end{eqnarray}

To compute the second term in Equation (\ref{ssexpMMin}), 
%our initial step is to 
we first consider the probability distribution function $p(s_{1}^{j},\bm{\theta}_{1}|\pmb{s}_{0},\bm{\theta}_0)$. Since  $\bm{\theta}_t$ is constant over time, we will put our focus on $p(s_{1}^{j}|\pmb{s}_{0},\bm{\theta}_0)$. As shown in Section~\ref{sec:multi-scaleBioprocess}, we've already derived the state transition model %of the continuous-time stochastic process 
in Equation (\ref{EMapprox}). This model adopts a truncated normal distribution, which is motivated by the fact that the concentration of molecules, represented by $s_{1}^{j}$, cannot be negative in reality. Furthermore, the introduction of the truncated normal distribution is due to the absence of an inverse moment for the standard normal distribution. To address this issue, we will first review Theorem~\ref{approx_trunc} and then apply it to approximate the second term in Equation (\ref{ssexpMMin}).
\begin{theorem}[\cite{INV1976}]
\noindent \vspace{0.1in}
Suppose $\mu>0$ and $X$ is a random variable with density 
\begin{equation*}\label{truncateN}
\phi(X)=\left\{
\begin{array}{l}\frac{k}{\sqrt{2\pi}\sigma}e^{\frac{-(X-\mu)^2}{2\sigma^2}},X\geq a>0,\\
         0,X<a,
\end{array}
\right.
\end{equation*}
where $k$ is the normalization constant. Then for each $\hat{\sigma}=\frac{\sigma}{\mu}\leq\frac{1}{5}$ and any value of $a$ satisfying the inequality $\hat{\sigma}^{2}\leq\frac{a}{\mu}\leq\frac{1}{25}$, we have:
%\begin{equation*}
%\label{truncateNexp}
$    \mu \mbox{E}[X^{-1}]=I(\hat{\sigma})+e_{1}, \left|e_{1}\right|<8000\hat{\sigma}^{12}<3.3\times 10^{-5},
$
%\end{equation*}
where $I(\hat{\sigma})$ is given in terms of Dawson's Integral by the following equations:
\begin{eqnarray}
I(\hat{\sigma})=\frac{\sqrt{2}}{\hat{\sigma}}D\left(\frac{1}{\sqrt{2}\hat{\sigma}}\right),\quad D(x)=e^{-x^{2}}\int_{0}^{x}e^{t^{2}}dt.
\end{eqnarray}
\label{approx_trunc}
\end{theorem}

%given $\bm{\theta}_t$, 
Before applying Theorem \ref{approx_trunc}, we rewrite Equation~(\ref{ssexpMMin}) by changing the variable, 
\begin{equation*}
    \int p(s_{1}^{j}|\pmb{s}_{0},\bm{\theta}_0)\frac{1}{K_{m}^{j}+s_{1}^{j}}ds_{1}^{j}\overset{\hat{s}_{1}^{j}=s_{1}^{j}+K_{m}^{j}}{=}\int p(\hat{s}_{1}^{j}|\pmb{s}_{0},\bm{\theta}_0)\frac{1}{\hat{s}_{1}^{j}}d\hat{s}_{1}^{j}.
\end{equation*}
When $\pmb{s}_0 \gg \Delta t$, both conditions $\hat{\sigma}\leq\frac{1}{5}$ and $\hat{\sigma}^{2}\leq\frac{a}{\mu}\leq\frac{1}{25}$ could be satisfied.  Thus, Theorem \ref{approx_trunc} is applicable to the approximation of $\int p(s_{1}^{j}|\pmb{s}_{0},\bm{\theta}_0)\frac{1}{K_{m}^{j}+s_{1}^{j}}ds_{1}^{j}$, which means:
\begin{equation}\label{Dapprox}
\int p(s_{1}^{j}|\pmb{s}_{0},\bm{\theta}_0)\frac{1}{K_{m}^{j}+s_{1}^{j}}ds_{1}^{j}\approx\frac{I(\hat{\sigma}_{0}^{j})}{\mu_{0}^{j}},
\end{equation}
\begin{equation}
\mu_{0}^{j}=s_{0}^{j}+\sum\limits_{k=1}^{R}\bm{N_{j,k}}\pmb{v}_{k}(\pmb{s}_{0},\bm{\theta}_0)\Delta t+K_{m}^{j}, ~~~\sigma_{0}^{j}=\sqrt{\sum\limits_{k=1}^{R}\bm{N}_{j,k}^{2}\pmb{v}_{k}(\pmb{s}_{0},\bm{\theta}_0)\Delta t},
    ~~~ \hat{\sigma}_{0}^j=\frac{\sigma_{0}^{j}}{\mu_{0}^{j}}.
    \label{stats}
\end{equation}
Since Dawson's Integral does not have an analytic expression, we introduce an analytic approximation of Dawson's Integral \citep{Dawappro}:
%\begin{equation*}
$
    D(x)\approx\frac{1}{2x}+\frac{1}{4x^{3}}$
    %$,
    for $x>2.68.
    $
%\end{equation*}
Based on Filobello-Nino's results, the approximation above can control the relative error below 2.5\% \citep{Dawappro}. Since we follow Theorem~\ref{approx_trunc}'s assumption that $\hat{\sigma}_{0}\leq \frac{1}{5}$, we could derive $\frac{1}{\sqrt{2}\hat{\sigma}_{0}}\geq\frac{1}{\frac{\sqrt{2}}{5}}>2.68$. Thus:
\begin{equation}\label{resclean}
\frac{I(\hat{\sigma}_{0}^{j})}{\mu_{0}^{j}}=\frac{\frac{\sqrt{2}}{\hat{\sigma}_{0}^{j}}D\left(\frac{1}{\sqrt{2}\hat{\sigma}_{0}^{j}}\right)}{\mu_{0}^{j}}\approx\frac{\frac{\sqrt{2}}{\hat{\sigma}_{0}^{j}}\frac{\sqrt{2}\hat{\sigma}_{0}^{j}}{2}\left(1+\frac{(\sqrt{2}\hat{\sigma}_{0}^{j})^2}{2}\right)}{\mu_{0}^{j}}=\frac{1+{(\hat{\sigma}_{0}^{j})^2}}{\mu_{0}^{j}}=\frac{{(\sigma_{0}^{j})^2}+{(\mu_{0}^{j})^2}}{{(\mu_{0}^{j})^3}}.
\end{equation}

Combining results from Equations (\ref{ssexp}), (\ref{ssexpMMin}), (\ref{Dapprox}) and (\ref{resclean}), we can get,
\begin{equation}
\mbox{E}(\pmb{s}_{2}|\pmb{s}_{0},\bm{\theta}_0) \approx f(\bm{v}(\pmb{s}_{0},\bm{\theta}_0))\triangleq \pmb{s}_{0}+\bm{N}\bm{v}(\pmb{s}_{0},\bm{\theta}_0)\Delta t+\bm{N}V_{max}\Delta t-\bm{N}\Delta t\left(V_{max}\odot K_{m}\odot \frac{I(\hat{\sigma}_{0})}{\mu_{0}}\right),\label{fsexp1}
\end{equation}
where 
$\frac{I(\hat{\sigma}_{0})}{\mu_{0}}$,$V_{max}$ and $K_{m}$ are $R$-dimensional vectors with $j$th component equal to $\frac{I(\hat{\sigma}_{0}^{j})}{\mu_{0}^{j}}$,$V_{max}^{j}$ and $K_m^j$.\\
In addition, we can derive $\mbox{Var}(\pmb{s}_2|{s}_{0},\bm{\theta}_0)$ following above process:
\begin{eqnarray} 
\mbox{Var}(\pmb{s}_{2}|\pmb{s}_{0},\bm{\theta}_0)&=&\int p(\pmb{s}_{1},\bm{\theta}_1|\pmb{s}_{0},\bm{\theta}_0)
\mbox{Var}(\pmb{s}_{2}|\pmb{s}_{1},\bm{\theta}_1)d(\pmb{s}_1,\bm{\theta}_1) 
\nonumber\\ 
&=&\int p(\pmb{s}_{1},\bm{\theta}_1|\pmb{s}_{0},\bm{\theta}_0)\left[\bm{N}diag(\bm{v}(\pmb{s}_{1},\bm{\theta}_1))\bm{N}^{\top}\Delta t\right]
d(\pmb{s}_1,\bm{\theta}_1) \nonumber\\ 
&=&\bm{N} diag\left(V_{max}-\left(V_{max}\odot K_{m}\odot \frac{I(\hat{\sigma}_{0})}{\mu_{0}}\right)\right)\bm{N}^{\top}\Delta t.\label{fsexp2}
\end{eqnarray}

Since we know the distribution of $\pmb{s}_2$ conditioned on $\pmb{s}_0$, we could extend it to time $t$ and write the form of $\mu$ and $\sigma$:
\begin{gather*}\label{SDE1}
d\pmb{s}_{t}=\mu(\pmb{s}_{t},\bm{\theta}_t)dt+\sigma(\pmb{s}_{t},\bm{\theta}_t)dW_t,\\ 
\mu(\pmb{s}_{t},\bm{\theta}_t)=
\mbox{E}[\pmb{s}_{t+2}-\pmb{s}_{t}|\pmb{s}_{t},\bm{\theta}_t]\approx\bm{N}\bm{v}(\pmb{s}_{t},\bm{\theta}_t)\Delta t+\bm{N}V_{max}\Delta t-\bm{N}\Delta t\left(V_{max}\odot K_{m}\odot \frac{I(\hat{\sigma}_t)}{\mu_t}\right),\label{adj3}\\
\sigma^2(\pmb{s}_{t},\bm{\theta}_t)=\mbox{Var}[\pmb{s}_{t+2}|\pmb{s}_{t},\bm{\theta}_t]\approx\bm{N} diag\left(V_{max}-\left(V_{max}\odot K_{m}\odot \frac{I(\hat{\sigma}_t)}{\mu_t}\right)\right)\bm{N}^{\top}\Delta t.\label{adjsigma}
\end{gather*}
We could also calculate $\frac{\partial \mu(\pmb{s}_{t}, \bm{\theta}_t)}{\partial \pmb{s}_t}$, $\frac{\partial \mu(\pmb{s}_{t},\bm{\theta}_t)}{\partial \bm{\theta}_t}$, $\frac{\partial \sigma(\pmb{s}_{t},\bm{\theta}_t)}{\partial \pmb{s}_t}$ and $\frac{\partial \sigma(\pmb{s}_{t},\bm{\theta}_t)}{\partial \bm{\theta}_t}$ which will be the input of the algorithm in Section~\ref{sec:LS-multiScaleModel}. 

\section{Local Sensitivity Analysis 
%Macro-kinetic State Transition Modeling
}
\label{sec:LS-multiScaleModel}

\subsection{Convergence Analysis}
\label{subsec:SA}

We start from the gradient of the backward flow represented in Equation (\ref{backward_calc}),
%Apparently, 
\begin{eqnarray*}
\triangledown\psi_{0,t_2}(\pmb{s}_{t_2},\bm{\theta}_{t_2})
&=&\triangledown \pmb{s}_{t_2}-\triangledown\int_{0}^{t_2}(\mu(\psi_{t,t_2}(\pmb{s}_{t_2},\bm{\theta}_{t_2})),f(\psi_{t,t_2}(\pmb{s}_{t_2},\bm{\theta}_{t_2})))^{\top}dt\\
&& - \triangledown\int_{0}^{t_2}(\sigma(\psi_{t,t_2}(\pmb{s}_{t_2},\bm{\theta}_{t_2})),g(\psi_{t,t_2}(\pmb{s}_{t_2},\bm{\theta}_{t_2})))^{\top}\circ d\widetilde{W}_t.
\end{eqnarray*}
The gradient $\triangledown \pmb{s}_0$ is the identity matrix $\mathbb{I}$ with $d$ dimension, where $d$ is the sum of the dimension of bioprocess states $\pmb{s}_t$ and the parameter set $\bm{\theta}_t$. For the last two terms on the right side of equation, based on Proposition~2.4.3 and Theorem~3.4.3 from \cite{kunita2019stochastic}, we can switch the order of derivative and integral, i.e., %then the above equations will be:
\begin{eqnarray}
    \triangledown\psi_{0,t_2}(\pmb{s}_{t_2},\bm{\theta}_{t_2})
    &=&\mathbb{I}_{d}-\int_{0}^{t_2}(\triangledown_\psi\mu(\psi_{t,t_2}(\pmb{s}_{t_2},\bm{\theta}_{t_2})),\triangledown_\psi f(\psi_{t,t_2}(\pmb{s}_{t_2},\bm{\theta}_{t_2})))^{\top}\triangledown\psi_{t,t_2}(\pmb{s}_{t_2},\bm{\theta}_{t_2})dt\nonumber\\
    && -\int_{0}^{t_2}(\triangledown_\psi\sigma(\psi_{t,t_2}(\pmb{s}_{t_2},\bm{\theta}_{t_2})),\triangledown_\psi g(\psi_{t,t_2}(\pmb{s}_{t_2},\bm{\theta}_{t_2})))^{\top}\triangledown\psi_{t,t_2}(\pmb{s}_{t_2},\bm{\theta}_{t_2})\circ d\widetilde{W}_t.\label{gradbflow}
\end{eqnarray}

Since $\psi_{0,t_2}$ is the inverse function of $\Phi_{0,t_2}$, we have $\Phi_{0,t_2}(\psi_{0,t_2}(\pmb{s}_{t_2},\bm{\theta}_{t_2}))=(\pmb{s}_{t_2},\bm{\theta}_{t_2})$. Therefore, by applying the chain rule, $\triangledown\Phi_{0,t_2}(\psi_{0,t_2}(\pmb{s}_{t_2},\bm{\theta}_{t_2}))\triangledown\psi_{0,t_2}(\pmb{s}_{t_2},\bm{\theta}_{t_2})=\mathbb{I}_{d}$. Then by applying Stratonovich version of Itô’s formula (Theorem 2.4.1 \citep{kunita2019stochastic}) on Equation (\ref{gradbflow}), we have,
\begin{equation*}
    \frac{\partial \triangledown\Phi_{0,t_2}(\psi_{0,t_2}(\pmb{s}_{t_2},\bm{\theta}_{t_2}))}{\partial \triangledown\psi_{0,t_2}(\pmb{s}_{t_2},\bm{\theta}_{t_2})}=\frac{\partial (\triangledown\psi_{0,t_2}(\pmb{s}_{t_2},\bm{\theta}_{t_2}))^{-1}}{\partial \triangledown\psi_{0,t_2}(\pmb{s}_{t_2},\bm{\theta}_{t_2})}=-\frac{1}{\triangledown\psi_{0,t_2}(\pmb{s}_{t_2},\bm{\theta}_{t_2})^2},
\end{equation*}
where
\begin{eqnarray}
\lefteqn{\triangledown\Phi_{0,t_2}(\psi_{0,t_2}(\pmb{s}_{t_2},\bm{\theta}_{t_2})) }
\nonumber \\
&=&\mathbb{I}_{d}-\int_{0}^{t_2}(\triangledown_\psi\mu(\psi_{t,t_2}(\pmb{s}_{t_2},\bm{\theta}_{t_2})),\triangledown_\psi f(\psi_{t,t_2}(\pmb{s}_{t_2},\bm{\theta}_{t_2})))^{\top}\triangledown\psi_{t,t_2}(\pmb{s}_{t_2},\bm{\theta}_{t_2}) %*
    \frac{\partial \triangledown\Phi_{t,t_2}(\psi_{t,t_2}(\pmb{s}_{t_2},\bm{\theta}_{t_2}))}{\partial \triangledown\psi_{t,t_2}(\pmb{s}_{t_2},\bm{\theta}_{t_2})}dt\nonumber\\
   & &-\int_{0}^{t_2}(\triangledown_\psi\sigma(\psi_{t,t_2}(\pmb{s}_{t_2},\bm{\theta}_{t_2})),\triangledown_\psi g(\psi_{t,t_2}(\pmb{s}_{t_2},\bm{\theta}_{t_2})))^{\top}\triangledown\psi_{t,t_2}(\pmb{s}_{t_2},\bm{\theta}_{t_2})
    %*
    \frac{\partial \triangledown\Phi_{t,t_2}(\psi_{t,t_2}(\pmb{s}_{t_2},\bm{\theta}_{t_2}))}{\partial \triangledown\psi_{t,t_2}(\pmb{s}_{t_2},\bm{\theta}_{t_2})}\circ d\widetilde{W}_t\nonumber\\
    &=&\mathbb{I}_{d}+\int_{0}^{t_2}(\triangledown_\psi\mu(\psi_{t,t_2}(\pmb{s}_{t_2},\bm{\theta}_{t_2})),\triangledown_\psi f(\psi_{t,t_2}(\pmb{s}_{t_2},\bm{\theta}_{t_2})))^{\top}\triangledown\Phi_{t,t_2}(\psi_{t,t_2}(\pmb{s}_{t_2},\bm{\theta}_{t_2}))dt\nonumber\\
    &&+\int_{0}^{t_2}(\triangledown_\psi\sigma(\psi_{t,t_2}(\pmb{s}_{t_2},\bm{\theta}_{t_2})),\triangledown_\psi g(\psi_{t,t_2}(\pmb{s}_{t_2},\bm{\theta}_{t_2})))^{\top}\triangledown\Phi_{t,t_2}(\psi_{t,t_2}(\pmb{s}_{t_2},\bm{\theta}_{t_2}))\circ d\widetilde{W}_t.\label{gradbflow_k}
\end{eqnarray}
%From $\triangledown\Phi_{0,t_2}(\psi_{0,t_2}(\pmb{s}_{t_2},\bm{\theta}_{t_2}))$, 
Let $\widetilde{A}_{0,t_2}(\pmb{s}_{t_2},\bm{\theta}_{t_2})=A_{0,t_2}(\psi_{0,t_2}(\pmb{s}_{t_2},\bm{\theta}_{t_2}))$.
Then, by combining Equation (\ref{gradbflow_k}) and 
\begin{equation*}
A_{0,t_2}(\pmb{s}_0,\bm{\theta}_0)=\triangledown \mathcal{L}(\Phi_{0,t_2}(\pmb{s}_0,\bm{\theta}_0))\triangledown\Phi_{0,t_2}(\pmb{s}_0,\bm{\theta}_0),\\
\end{equation*}
we can derive
\begin{eqnarray}
\lefteqn{\widetilde{A}_{0,t_2}(\pmb{s}_{t_2},\bm{\theta}_{t_2})
=A_{0,t_2}(\psi_{0,t_2}(\pmb{s}_{t_2},\bm{\theta}_{t_2}))=\triangledown \mathcal{L}(\Phi_{0,t_2}(\psi_{0,t_2}(\pmb{s}_{t_2},\bm{\theta}_{t_2})))\triangledown\Phi_{0,t_2}(\psi_{0,t_2}(\pmb{s}_{t_2},\bm{\theta}_{t_2}))}
\nonumber\\
&=&\triangledown\mathcal{L}(\pmb{s}_{t_2})+\int_{0}^{t_2}( \triangledown_\psi\mu(\psi_{t,t_2}(\pmb{s}_{t_2},\bm{\theta}_{t_2})),\triangledown_\psi f(\psi_{t,t_2}(\pmb{s}_{t_2},\bm{\theta}_{t_2})))^{\top}\widetilde{A}_{t,t_2}(\pmb{s}_{t_2},\bm{\theta}_{t_2})dt\nonumber\\
&&+\int_{0}^{t_2}(\triangledown_\psi\sigma(\psi_{t,t_2}(\pmb{s}_{t_2},\bm{\theta}_{t_2})),\triangledown_\psi g(\psi_{t,t_2}(\pmb{s}_{t_2},\bm{\theta}_{t_2})))^{\top}\widetilde{A}_{t,t_2}(\pmb{s}_{t_2},\bm{\theta}_{t_2})\circ d\widetilde{W}_t.\label{gradbflow_wA}
\end{eqnarray}

From Equations (\ref{backward_calc}) and (\ref{gradbflow_wA}), we can see that the drift and diffusion terms for $\psi_{0,t_2}(\pmb{s}_{t_2},\bm{\theta}_{t_2})$ and $\widetilde{A}_{0,t_2}(\pmb{s}_{t_2},\bm{\theta}_{t_2})$ are $C_b^{\infty,1}$, which means the system $(\widetilde{A}_{0,t_2}(\pmb{s}_{t_2},\bm{\theta}_{t_2}),\psi_{0,t_2}(\pmb{s}_{t_2},\bm{\theta}_{t_2}))$ has a unique strong solution. Thus, we could define $\widetilde{A}_{0,T}(\pmb{s}_{t_2},\bm{\theta}_{t_2})=F(\pmb{s}_{t_2},\bm{\theta}_{t_2},W)$ where $W=\{W_t\}_{0\leq t \leq T}$ is a path of Wiener process and $F: \mathbb{R}^d \times C([0,1],\mathbb{R}^{J}) \to \mathbb{R}^d$ is a deterministic measurable function, which is also called Itô map. In our case, the number of reactions denoted by $J$ defines the dimension of the Wiener process deployed to simulate the stochastic reaction network.
Note that $A_{0,t_2}(\pmb{s}_0,\bm{\theta}_0)=\widetilde{A}_{0,t_2}(\Phi_{0,t_2}(\pmb{s}_0,\bm{\theta}_0))$. In Section~{\ref{subsec:MU&SA}}, we've already shown that $\Phi_{0,t_2}(\pmb{s}_0,\bm{\theta}_0)$ has a unique strong solution. Similar to $F$, we define $G: \mathbb{R}^d \times C([0,1],\mathbb{R}^{J}) \to \mathbb{R}^d$ as the solution map for forward flow, i.e., $G(\pmb{s}_0,\bm{\theta}_0,W)=\Phi_{0,T}(\pmb{s}_0,\bm{\theta}_0)$. Then, apparently:
\begin{equation*}
    A_{0,T}(\pmb{s}_0,\bm{\theta}_0)=\widetilde{A}_{0,T}(G(\pmb{s}_0,\bm{\theta}_0,W))=F(G(\pmb{s}_0,\bm{\theta}_0,W),W).
\end{equation*}

\begin{comment}
We could follow numerical schemes to develop solvers $F_h$ and $G_h$ for $F$ and $G$, where $h=\frac{T}{L}$ denotes the size of the grid for solvers. Then, based on Theorem 3.3 in \cite{Li2020ScalableGF}, we have,
\begin{equation}
    F_h(G_h(\pmb{s}_0,\bm{\theta}_0,W),W)\to F(G(\pmb{s}_0,\bm{\theta}_0,W),W)=A_{0,T}(\pmb{s}_0,\bm{\theta}_0), \forall \pmb{s}_0 \in \mathbb{R}^{d}.\label{algo_conv}
\end{equation}
\end{comment}

\subsection{Algorithm Development}
\label{sec:algorithm}
Based on Equations (\ref{forward_calc}), (\ref{backward_calc}), and (\ref{gradbflow_wA}), with $\frac{\partial \mu(\pmb{s}_{t},\bm{\theta}_t)}{\partial \pmb{s}_t}$, $\frac{\partial \mu(\pmb{s}_{t},\bm{\theta}_t)}{\partial \bm{\theta}_t}$, $\frac{\partial \sigma(\pmb{s}_{t},\bm{\theta}_t)}{\partial \pmb{s}_t}$ and $\frac{\partial \sigma(\pmb{s}_{t},\bm{\theta}_t)}{\partial \bm{\theta}_t}$ as inputs, we could approximate $A_{0,T}(\pmb{s}_0,\bm{\theta}_0)$. Since the system is described in Stratonovich integral, the most commonly used Euler-Maruyama Scheme is not applicable, which requires the system to be represented in Itô integral. 
Thus, we instead deployed the Euler-Heun method below:
\begin{eqnarray*}
    (\pmb{s}_{t+1},\pmb{\theta}_{t+1})^{\top}&=&(\pmb{s}_{t},\pmb{\theta}_{t})^{\top}+(\mu(\pmb{s}_t,\bm{\theta}_t),f(\pmb{s}_t,\bm{\theta}_t))^{\top}\Delta t
    \nonumber \\
    && +\frac{1}{2}\left[(\sigma(\pmb{s}_t,\bm{\theta}_t),g(\pmb{s}_t,\bm{\theta}_t))^{\top}+(\sigma(\overline{\pmb{s}}_t,\overline{\bm{\theta}}_t),g(\overline{\pmb{s}}_t,\overline{\bm{\theta}}_t))^{\top}\right](W_t-W_{t-1}),\\
(\overline{\pmb{s}}_t,\overline{\bm{\theta}}_t)^{\top}&=&(\pmb{s}_t,\bm{\theta}_t)^{\top}+(\sigma(\pmb{s}_t,\bm{\theta}_t),g(\pmb{s}_t,\bm{\theta}_t))^{\top}(W_t-W_{t-1}).
\end{eqnarray*}

Based on the schemes, the procedure of adjoint sensitivity analysis on the SDEs-based mechanistic model is described in Algorithm~\ref{algo_adjoint_sens}. {The algorithm is built on the Euler-Heun Scheme, using the gradients of the drift and diffusion terms as inputs, and produces the expected gradient of the simulation prediction MSE as its output.}
The algorithm begins with the generation of sample paths of the Wiener process $W_t^n$ in Step~1. Subsequently, as depicted in Steps 2 and 3, we conduct simulations under $\pmb{s}_0$ and $\bm{\theta}_0$ to obtain the simulated output denoted by $O(\pmb{s})$. In the empirical study, the observed states $\bm{s}^c$ are generated from the model with a predefined set of true parameters $\bm{\theta}^{c}$, using the same path $W_t^n$ to control randomness. $W_t^n$ is queried again in the backward pass $\widetilde{W}_t^{n}$ to generate the backward flow $\psi^n_{(t-1)\Delta t,T}(\Phi_{0,T}^{n}(\pmb{s}_0,\bm{\theta}_0))$. By combining observed output, simulated output, and the backward flow, we derive sensitivity $\widetilde{A}^n_{(t-1)\Delta t,T}(\Phi_{0,T}^n(\pmb{s}_0,\bm{\theta}_0))$ corresponding to path $W_t^n$ in Step 6. 
By repeating these steps, we generate $N$ sample paths, and the mean value of $\widetilde{A}^n_{(t-1)\Delta t,T}(\Phi_{0,T}^n(\pmb{s}_0,\bm{\theta}_0))$ for each sample path serves as the final sensitivity, as shown in Step 7.  
%please (1) add the label of algorithm, say; (2) add the description of this algorithm; (3) update the notation in the algorithm; and (4) add the equation index for the associated steps }

\begin{algorithm}[th] 
\DontPrintSemicolon
\KwIn{%$t$, timestep;

\begin{itemize}
  \item Observed states $\pmb{s}^{c}$ from the real system.
  \item Initial parameters $\bm{\theta}_{0}$.
  \item Start time $t_0$, end time $T$.
  \item Gradient of drift and diffusion term $\frac{\partial \mu(\pmb{s}_{t},\bm{\theta}_t)}{\partial \pmb{s}_t}$, $\frac{\partial \mu(\pmb{s}_{t},\bm{\theta}_t)}{\partial \bm{\theta}_t}$, $\frac{\partial \sigma(\pmb{s}_{t},\bm{\theta}_t)}{\partial \pmb{s}_t}$ and $\frac{\partial \sigma(\pmb{s}_{t},\bm{\theta}_t)}{\partial \bm{\theta}_t}$.
  \item Number of iterations $N$.
  \item Grid size $\Delta t$, number of steps $S=\frac{T}{\Delta t}$.
\end{itemize}
%$\bm{\epsilon} = \{\epsilon_1,\cdots,\epsilon_G\}$, the value of distance tolerance bound; %$M$, the number of nearest neighbours; %$\{(\pmb{\theta}_{t,i}^{(w-1)})_{i=1}^N, (q_{t,i}^{(w-1)})_{i=1}^N\}$ %$\{\theta_t^{w-1}, q_t^{w-1}\}$ 
%the posterior distribution samples of previous Gibbs iteration and the corresponding weight, if exist;
}
\KwOut{$\mbox{E}\left[\left. \left(\frac{\partial (O(\pmb{s})-O(\pmb{s}^c))^2}{\partial \pmb{s}_0},\frac{\partial (O(\pmb{s})-O(\pmb{s}^c))^2}{\partial \bm{\theta}_0}\right)^{\top} \right| (\pmb{s}_0,\bm{\theta}_0)\right]$
} 
\textbf{1.} Calculate the real system output $O(\pmb{s}^{c})$ and define $\pmb{s}_{0}=\pmb{s}_{0}^{c}$;\\
{   
    \For{$n = 1,\ldots, N$}{
    \For{$t \gets 1$ \KwTo $S$}{
    \textbf{2.} Generate a sample path of Wiener process $W^n_t$ with grid size $\Delta t$; \\
    \textbf{3.} Based on $W^n_t$, generate $\Phi^n_{0,t\Delta t}(\pmb{s}_0,\bm{\theta}_0)$ from Equation (\ref{forward_calc}) with Euler-Heun Scheme; \\
    }
    \textbf{4.} Generate backward pass $\widetilde{W}^n_t$ based on forward pass $W_t^n$;\\
    \textbf{5.} Calculate the model prediction output $O(\Phi^n_{0,T}(\pmb{s}_0,\bm{\theta}_0))$;\\

    \For{$t \gets S$ \KwTo $1$}{
    \textbf{6.} Based on $\widetilde{W}^n_t$, generate $\psi^n_{(t-1)\Delta t,T}(\Phi_{0,T}^{n}(\pmb{s}_0,\bm{\theta}_0))$ from Equation (\ref{backward_calc}) with Euler-Heun Scheme;\\
    \textbf{7.} Based on $\psi^n_{(t-1)\Delta t,T}(\Phi_{0,T}^{n}(\pmb{s}_0,\bm{\theta}_0))$, generate $\widetilde{A}^n_{(t-1)\Delta t,T}(\Phi_{0,T}^n(\pmb{s}_0,\bm{\theta}_0))$ from Equation (\ref{gradbflow_wA}) with Euler-Heun Scheme, $\triangledown\mathcal{L}(\pmb{s}_{T})=2|O(\pmb{s})-O(\pmb{s}^{c})|$;}
    
    }
\textbf{8.} Calculate $
\mbox{E}\left[\left. \left(\frac{\partial (O(\pmb{s})-O(\pmb{s}^c))^2}{\partial \pmb{s}_0},\frac{\partial (O(\pmb{s})-O(\pmb{s}^c))^2}{\partial \bm{\theta}_0}\right)^{\top} \right| (\pmb{s}_0,\bm{\theta}_0)\right]=\frac{1}{N}\sum\limits_{n=1}^N\widetilde{A}^{n}_{0,T}(\Phi_{0,T}^n(\pmb{s}_0,\bm{\theta}_0))$
}
\caption{Adjoint Sensitivity Analysis on SDEs based on the MSE of Output Prediction.}
\label{algo_adjoint_sens}
\end{algorithm}

\noindent 

From Theorem \ref{algo_conv}, as the grid or time size $\Delta t \to 0$, the output of Algorithm \ref{algo_adjoint_sens} will converge pathwise (i.e., almost surely) from any fixed starting point to true local sensitivity. This sensitivity analysis, accounting for the interdependency of multi-scale mechanistic model, accelerates 
%the calibration of $\pmb{\theta}$ to achieve our further objective, i.e., $\pmb{\theta}$ that 
the search for the optimal parameters $\pmb{\theta}$ that
minimizes the model prediction's MSE.
\begin{theorem}[\cite{Li2020ScalableGF}]\label{algo_conv}
Suppose the schemes $F_{h}$ and $G_{h}$, with $h=\frac{T}{L}$ denoting the size of the grid for solvers, satisfy the following conditions: 
\begin{itemize}
    \item $F_{h}(\pmb{s}_t,\pmb{\theta}_t, W) \to F(\pmb{s}_t,\pmb{\theta}_t, W)$, $G_{h}(\pmb{s}_t,\pmb{\theta}_t, W) \to G(\pmb{s}_t,\pmb{\theta}_t, W)$ as $h \to 0$;
    \item For $\forall M>0$, $\sup_{\pmb{s}_t;\pmb{\theta}_t\leq M} |F_{h}(\pmb{s}_t,\pmb{\theta}_t, W)- F(\pmb{s}_t,\pmb{\theta}_t, W)| \to 0$ as $h \to 0$.
\end{itemize}
Then for any starting point $\pmb{s}_0$ and $\pmb{\theta}_0$, we have,
\begin{equation}
    F_h(G_h(\pmb{s}_0,\bm{\theta}_0,W),W)\to F(G(\pmb{s}_0,\bm{\theta}_0,W),W)=A_{0,T}(\pmb{s}_0,\bm{\theta}_0), \forall \pmb{s}_0 \in \mathbb{R}^{d}.
\end{equation}
\end{theorem}

\begin{comment}
The algorithm begins with the generation of sample paths of the Wiener process $W_t^n$ in Step~1. Subsequently, as depicted in Steps 2 and 3, we conduct simulations under $\pmb{s}_0$ and $\bm{\theta}_0$ to obtain the simulated output $O(\pmb{s})$. In the empirical study, the observed states $\bm{s}^c$ are generated from the model with a predefined set of true parameters $\bm{\theta}^{c}$, using the same path $W_t^n$ to control randomness. $W_t^n$ is queried again in the backward pass $\widetilde{W}_t^{n}$ to generate the backward flow $\psi^n_{(t-1)\Delta t,T}(\Phi_{0,T}^{n}(\pmb{s}_0,\bm{\theta}_0))$. By combining observed output, simulated output, and the backward flow, we derive sensitivity $\widetilde{A}^n_{(t-1)\Delta t,T}(\Phi_{0,T}^n(\pmb{s}_0,\bm{\theta}_0))$ corresponding to path $W_t^n$ in Step 6. In total, we generate $N$ sample paths, and the mean value of $\widetilde{A}^n_{(t-1)\Delta t,T}(\Phi_{0,T}^n(\pmb{s}_0,\bm{\theta}_0))$ for each sample path serves as the final sensitivity, as shown in Step 7.  
\end{comment}

\section{Empirical Study}
\label{sec: empirical study}

In this section, we conduct the empirical study by using the In-Vitro Transcription (IVT) system example to validate our SA algorithm. %IVT is a procedure enabling the template-directed synthesis of RNA molecules of any sequence and the yield $\pmb{Y}$ of the system is mRNA, which is included in bioprocess states $\pmb{s}$ of the system, as yield $\pmb{Y}$. For inputs to Algorithm {\ref{algo_adjoint_sens}}, we define the original parameters from \cite{wangIVT} as $\bm{\theta}_c$. 
The bioprocess model from \cite{wangIVT} with the true model parameters denoted by $\bm{\theta}^c$ is used to assess the performance of proposed adjoint SA algorithm.
%Then, we add a random perturbation to $\bm{\theta}_c$ to generate $\bm{\theta}_0$. 
We start the section by providing an in-depth interpretation of the sensitivity analysis results obtained from Algorithm \ref{algo_adjoint_sens}. 
%Then, guided by these findings, we proceed to manually adjust $\bm{\theta}_0$ using a study rate $\alpha=0.01$. 
We then assess the efficacy of our approach by showing the decrease rate in the simulation output prediction MSE over multiple search iterations of model parameters.

\begin{figure}
\centering
\subfloat[]{\includegraphics[width=0.5\textwidth]{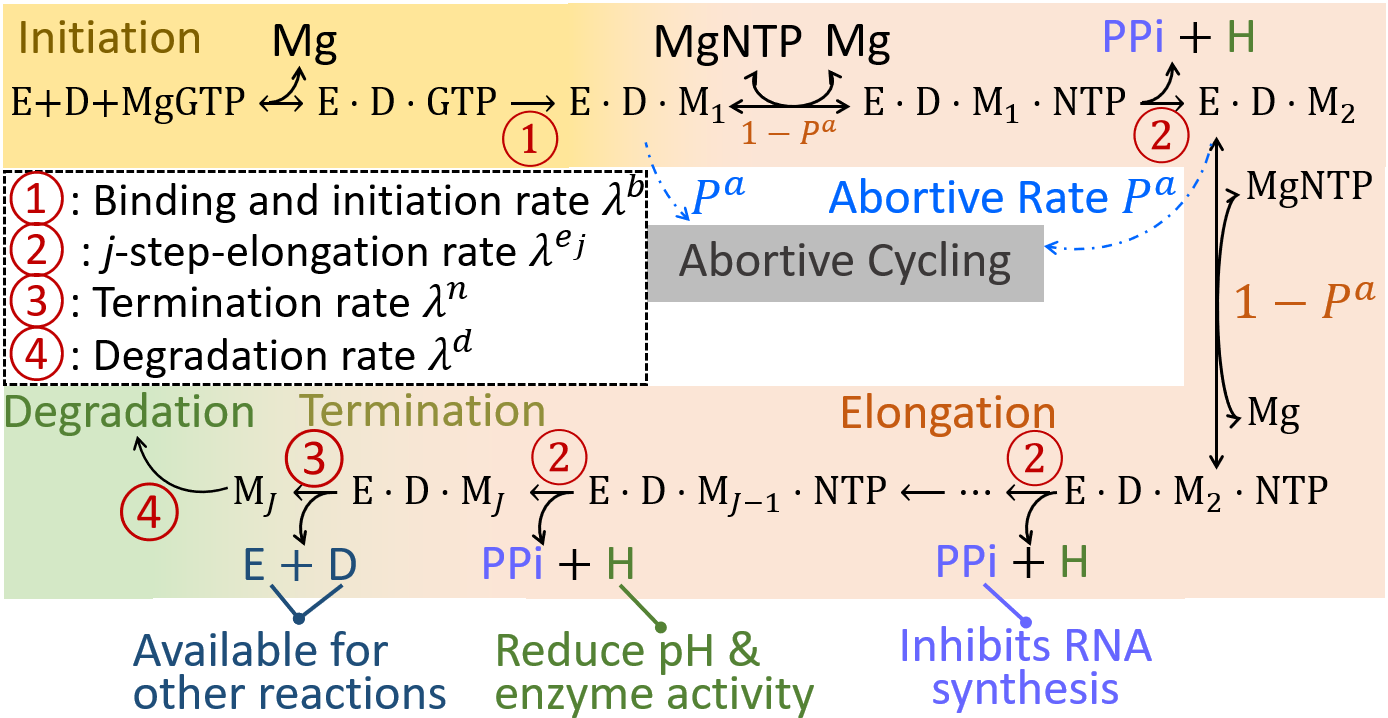}}
\subfloat[]{\includegraphics[width=0.3\textwidth]{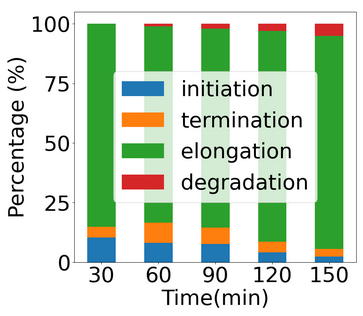}}
\caption{Interaction between subprocesses of the IVT system and their relative importance. (a) The reaction network for the IVT process \protect\citep{wangIVT}. (b) Relative importance of the IVT system 
 subprocesses parameter estimation on the output predictions at different times based on SA results.}
\label{IVT_inte_rela}
\end{figure}

The IVT process is mainly composed of four sub-processes: (1) initiation; %, in which enzyme binds to the promoter region of DNA template to form the transcription initiation complex; 
(2) elongation; %, in which a stable and processive enzyme elongation complex is formed; 
(3) termination; %, in which the interaction between RNAP and DNA template is disrupted upon encountering a terminator sequence or signal; 
and (4) degradation of 
%the synthesized full-length 
generated RNA molecule product. 
The molecular reaction network is illustrated in Figure~\ref{IVT_inte_rela}.
One insightful interpretation of the sensitivity analysis results is to investigate the relative importance of each subprocess, which is inferred from the sensitivity of their corresponding parameters. Aligned with this reasoning, Figure \ref{IVT_inte_rela} presents the relative importance of each subprocess within the IVT system. Notably, the result underscores the pivotal role of elongation reactions in the IVT process regardless of termination time $T$. Furthermore, it is observed that the impact of degradation increases over time, which fits our expectation since the synthesis rate of mRNA transcripts decreases as the IVT process approaches completion. Concurrently, the accumulation of generated RNA transcripts leads to an increase in the degradation rate at the system level. Therefore, the relevance of parameters associated with the degradation model becomes increasingly significant over time. Conversely, the influence of the initiation process appears to diminish over time as the raw materials (i.e., NTPs) used to synthesize RNA products are consumed.

\begin{comment}
\begin{figure}
\centering
\begin{subfigure}{.5\textwidth}
  \centering
  \includegraphics[width=\linewidth]{exp30.png}
  \caption{$T=30$}
  \label{fig:sub1}
\end{subfigure}%
\begin{subfigure}{.5\textwidth}
  \centering
  \includegraphics[width=\linewidth]{exp60.png}
  \caption{$T=60$}
  \label{fig:sub2}
\end{subfigure}
\label{fig:test}
\bigskip
\begin{subfigure}{.5\textwidth}
  \centering
  \includegraphics[width=\linewidth]{exp90.png}
  \caption{$T=90$}
  \label{fig:sub3}
\end{subfigure}%
\begin{subfigure}{.5\textwidth}
  \centering
  \includegraphics[width=\linewidth]{exp.png}
  \caption{$T=150$}
  \label{fig:sub4}
\end{subfigure}
\caption{Comparison of parameter sensitivities.}
\label{compps}
\end{figure}
\end{comment}

To further assess the performance of the proposed adjoint SA, we benchmark our approach with a state-of-art gradient estimation approach \citep{Fu2015} that estimates the gradient by using finite difference (FD) simulation estimator, i.e., %, the gradient estimation approach could be divided into two main categories — indirect and direct. Our approach utilizes structural information inside the reaction network and should be classified as a direct gradient estimation approach. For comparison, we select an indirect gradient estimator that uses only final yields $\pmb{Y}_T$ for the gradient estimation as the benchmark. Specifically, the form of the estimator is:
\begin{equation*}
\frac{O(\Phi^n_{0,T}(\pmb{s}_0,\hat{\bm{\theta}}+c_k\pmb{e}_k))-O(\Phi^n_{0,T}(\pmb{s}_0,\hat{\bm{\theta}}-c_k\pmb{e}_k))}{2c_k}
\end{equation*}
for the $n$-th iteration of parameter search, where $c_k$ is the perturbation and $\pmb{e}_k$ denotes a unit vector with the element in $k$-th dimension equal to 1 and all remaining elements equal to 0. In our case, we define $c_k=p\times \hat{\bm{\theta}}_k$ where $p$ is a percentage selected to be $5\%$, $10\%$, and $20\%$. 
%We choose multiple percentiles here since $c_k$ influences the accuracy of the gradient estimator. % so we need to determine appropriate $c_k$.

\begin{table}[h]
    \begin{center}

    \begin{tabular}{ | >{\centering\arraybackslash}m{0.6in} | 
    >{\centering\arraybackslash}m{0.8in} |
    >{\centering\arraybackslash}m{0.8in} | 
    >{\centering\arraybackslash}m{0.8in} |
    >{\centering\arraybackslash}m{0.8in} |
    >{\centering\arraybackslash}m{0.8in} |}

    \hline

    \textbf{Approach}& Observed& Adjoint sensitivity& Finite difference (p=$5\%$) & Finite difference (p=$10\%$)  & Finite difference (p=$20\%$)\\[0ex] \hline
    \textbf{Yield ($\mu M$)}& $73.48\pm0.28$ & $73.05\pm0.27$ & $72.66\pm0.32$ &  $74.37\pm0.39$ & $74.51\pm0.31$ \\[0ex] \hline

  \end{tabular}
  \end{center}
\caption{Final RNA yield and its 95$\%$ confidence interval with $\hat{\bm{\theta}}$ calibrated with different approaches.}
\label{yields}
\end{table}

\begin{figure}[h]
\centering
\subfloat[MSE]{\includegraphics[width=0.4\textwidth]{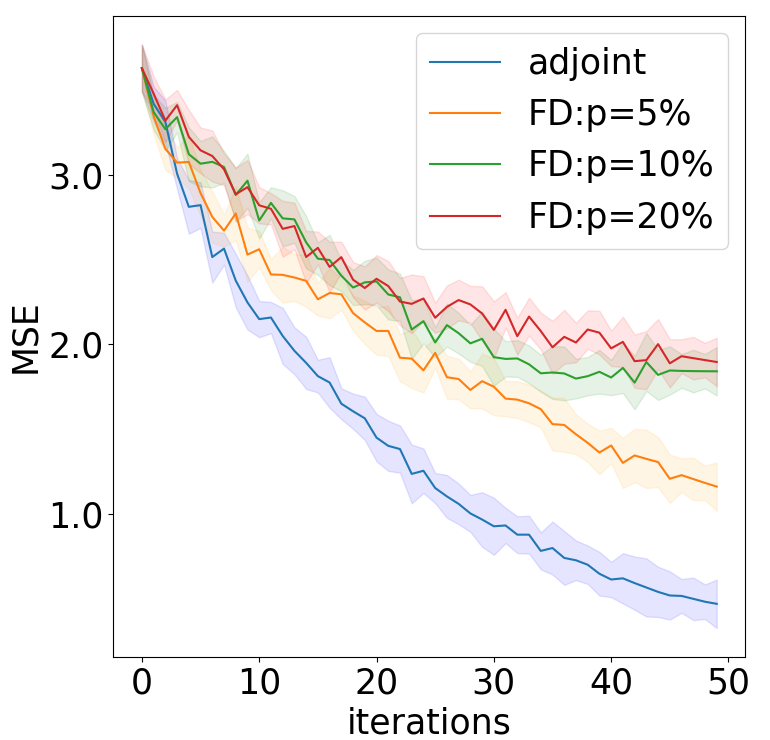}}\hskip1ex
\subfloat[Relative error between $\hat{\bm{\theta}}$ and $\bm{\theta}^c$]{\includegraphics[width=0.4\textwidth]{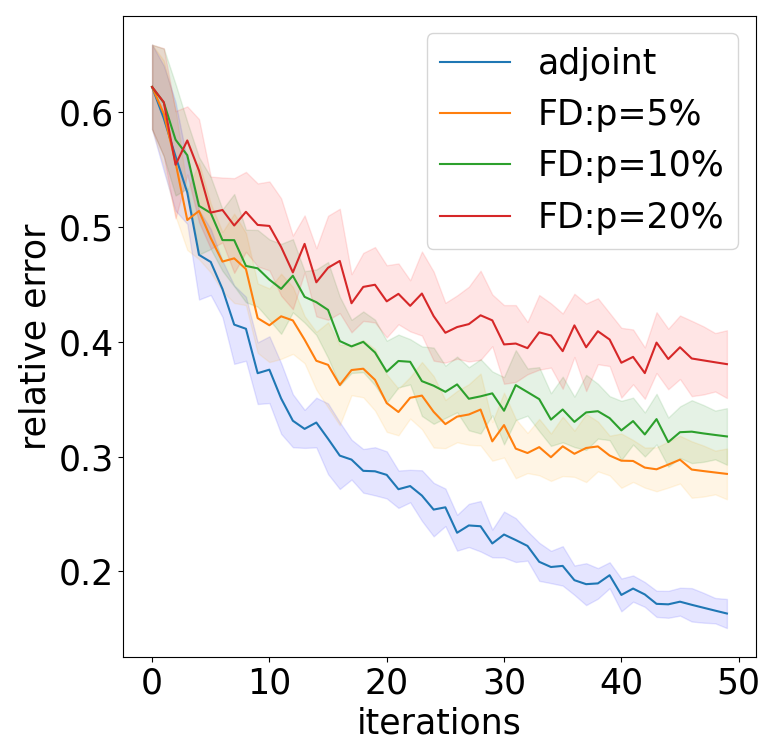}}
\caption{The results of IVT output prediction MSE and model parameter estimation relative error obtained by using adjoint SA and finite difference approaches.}
\label{Measure_grad}
\end{figure}

Figure \ref{Measure_grad} illustrates the convergence of the Mean Squared Error (MSE) of IVT process output prediction over the iterations. It's evident that as the number of iterations increases, the adjoint SA approach surpasses the finite difference gradient estimation approach.
%with a smaller lower bound of error no matter which percentile we select for the finite difference approach. 
Nonetheless, same as the finite difference estimator, the convergence speed gradually decreases, and the lower bound shows no further improvement once it reaches approximately $0.4 (g/L)^2$, which could be associated with a fixed step size, i.e., $\alpha=0.01$, used during the gradient search process.
%$\pmb{\hat{\theta}}_{n}=\pmb{\hat{\theta}}_{n-1}+\alpha*\mbox{E}\left[ \frac{\partial (O(\pmb{s}(\bm{\hat{\theta}}_{n-1}))-O(\pmb{s}^c))^2}{\partial \bm{\hat{\theta}}_{n-1}} \right]$. $\pmb{s}(\bm{\hat{\theta}}_{n-1})$ is the simulated state with $\bm{\hat{\theta}}_{n-1}$ as the input of simulation model.} 
This slowdown could also be attributed to our MSE objective measure that solely relies on the final RNA yield. This discrete data provide very limited information about the interactions within the complex reaction network. We also show the specific RNA yield prediction by using the model with parameters  $\hat{\bm{\theta}}$ estimated with different approaches in Table \ref{yields}.  In contrast, the final RNA yield and its $95\%$ confidence interval (CI) obtained from the system with true parameters $\bm{\theta}^c$ are $73.48\pm0.28$. {
There is an overlap between the 95\% confidence interval of the adjoint SA approach's predictions and the observed RNA yield, indicating no statistically significant difference between them. In contrast, the finite difference approach shows a significant prediction discrepancy from the observed values, demonstrating that the adjoint SA approach outperforms the finite difference method.}
%which resemble the corresponding results of our approach.

Furthermore, we present the results of the expected relative error between $\hat{\bm{\theta}}$ and $\bm{\theta}^c$, i.e., $\mbox{E}\big[|\frac{\hat{\bm{\theta}}-\bm{\theta}^c}{\bm{\theta}^c}|\big]$, in Figure \ref{Measure_grad}. 
Our approach consistently outperforms the finite difference estimator, as the expected relative error diminishes more significantly after 50 iterations compared with the finite difference estimator regardless of the $p$ value chosen. However, similar to the MSE results, the lower bound of the relative difference stagnates at approximately $15\%$, which could be attributed to the same limitation mentioned above. 

\section{CONCLUSION}
\label{sec: conclusion}

In this paper, we present an adjoint sensitivity analysis approach designed to expedite the search for stochastic reaction network model parameters .
We first formulate the multi-scale bioprocess stochastic reaction network in the form of SDEs. Then, we develop an adjoint SA algorithm on SDEs for computing local sensitivities of model parameters and validate the convergence of the algorithm. Our empirical study underscores the importance of model parameters of enzymatic stochastic reaction networks and provides empirical evidence of the efficacy of our approach. 
%In contrast to the indirect gradient estimator, our approach has the capability to incorporate measurements of intermediate states. 
Moving forward, we aim to leverage this capability by refining the loss function. Rather than solely relying on the Mean Squared Error (MSE) of the final output, we intend to integrate additional bioprocess states and consider the disparities between simulated and real trajectories, which promises to enhance the effectiveness and accuracy of our method.

% \section*{ACKNOWLEDGMENTS}

\bibliography{wscbib} 
\clearpage

\end{document}